\documentclass[12pt]{article}
\usepackage{amsmath,amssymb,bm}
\usepackage{gensymb}
\usepackage{nicefrac} 
\usepackage{epsfig}
\usepackage{graphicx}
\usepackage{slashed}
\usepackage{epsfig} \usepackage{graphicx} \usepackage{color}
\usepackage{mathrsfs} \usepackage{amssymb} \usepackage{amsmath} \usepackage{url}

\usepackage{xspace} 
\usepackage[bookmarks, breaklinks, colorlinks,urlcolor=black, citecolor=red, 
linkcolor=blue]{hyperref}
\usepackage{braket}
\usepackage[toc,page]{appendix}

\newcommand{\A}{\mathcal{A}}
\newcommand{\e}{\epsilon}

\def \PPbr{$P^0-\bar{P^0}$ \ }
\def \BBbr{$B^0-\bar{B^0}$ \ }

\def \oln{\bar}
\def \widebar{\bar}
\def \pl{\parallel}
\def \ls{{\lambda\sigma}}
\def \c2b{\cos 2\beta}
\def \s2b{\sin 2\beta}

\def\T {\ensuremath{T}\xspace}
\def\CP {\ensuremath{C\!P}\xspace}
\def\CPT {\ensuremath{C\!P T}\xspace}

\def\beq{\begin{equation}}
\def\eeq{\end{equation}}
\def\bea{\begin{eqnarray}}
\def\eea{\end{eqnarray}}
\def\nn{\nonumber}

\def\roughly#1{\mathrel{\raise.3ex\hbox
		{$#1$\kern-.75em\lower1ex\hbox{$\sim$}}}}

\allowdisplaybreaks
\begin{document}


\begin{center}
	\bigskip
	{\large \bf \boldmath Behaviour of observables for neutral meson decaying to two vectors in the presence of \T, \CP and \CPT violation in mixing only}		
	\\
	\bigskip
	\bigskip
	{\large
		Anirban Karan $^{a,}$\footnote{kanirban@iith.ac.in}, 
	    Abinash Kumar Nayak $^{b,}$\footnote{abinashkn@imsc.res.in}}
	
\end{center}

\begin{flushleft}
	~~~~~~~~~~~$a$: {\it Indian Institute of Technology, Hyderabad,}\\
	~~~~~~~~~~~~~~~{\it  Kandi, Sangareddy, Telangana 502285, India}\\
	~~~~~~~~~~~$b$: {\it The Institute of Mathematical Sciences,}\\
	~~~~~~~~~~~~~~~{\it Taramani, Chennai 600113, India;}\\
	~~~~~~~~~~~~~~~{\it Homi Bhabha National Institute,}\\
	~~~~~~~~~~~~~~~{\it Anushaktinagar, Mumbai 400094, India}
\end{flushleft}

\begin{center}
	\bigskip (\today)
	\vskip0.5cm {\Large Abstract\\} \vskip3truemm
	\parbox[t]{\textwidth}{	When a neutral meson $(P^0 \text{ or } \bar P^0)$ decays to two vector particles, a large number of observables can be constructed from differential decay rate based on the polarization of final state. But, theoretically, all of them are not independent to each other and hence, some relations among observables emerge. These relations have been well studied in the scenario with no \T and \CPT violation in neutral meson mixing and no direct \CP violation as well. In this paper, we have studied the relations among observables in the presence of \T, \CP and \CPT violating effects in mixing only. We find that except four of them, all the other old relations get violated and new relations appear if \T and \CPT violations in mixing are present. Invalidity of these relation will signify the presence of direct violation of \T, \CP and \CPT (i.e. violation in the decay itself).}
\end{center}


\thispagestyle{empty}
\newpage
\setcounter{page}{1}
\baselineskip=14pt

\section{Introduction:}
\label{Intro}

\CPT invariance is believed to be a sacred principle of any locally Lorentz invariant quantum field theory. In any axiomatic quantum field theory, this discrete symmetry emerges to be exact up to any order. It has a direct connection with the preservation of Lorentz symmetry \cite{Greenberg,CPTLorentz2}. Due to its great theoretical importance, it is necessary to test the validity of this principle experimentally. \CPT invariance predicts the masses or lifetimes of any particle and its anti-particle to be the same, which has been tested for lots of particles through direct experiments \cite{pdg}. But one can argue that these quantities are usually dominated by strong or electromagnetic interactions and hence there exits a possibility for tiny \CPT violating effects, mediated by weak interactions, to be undetectable in those direct experiments. In this regard, mixing of neutral pseudoscalar meson ($K^0,\, D^0,\, B_d^0,\, B_s^0$) with its own antiparticle is a promising area \cite{Lavora} to search for \CPT violating effects as this phenomenon is a second order electroweak process. However, since the most general mixing matrix includes \T and \CP violating parameters as well, we have to study the effects of \CP, \T and \CPT violation together.

Searches for \CP, \T and \CPT violation using leptonic and semi-leptonic channels as well as the modes where neutral pseudoscalar meson decays to two other pseudoscalars or one vector and one pseudoscalar have been performed extensively \cite{CPTVBmix_th2,Alvarez:2003kh,CPTVBmix_BaBar1,CPTVBmix_BaBar2,Alvarez:2004tj,CPTVBmix_th3,CPTVBmix_th4,expts1,CPTVBmix_th5,kundu1,kundu2,tilburg,Bernabeu:2016sgz,expts2,CPTps,Botela,hflav}. However, effects of \CPT violation on the modes where neutral pseudoscalar meson decays to two vectors $(P^0 \text{ or } \bar P^0\to V_1 V_2)$ are not very well studied. Though the Refs. \cite{VV1,VV2,London:2003rk} discuss about these modes involving two vectors, they only consider the SM scenario (i.e. only \CP violation in mixing) and its extension to a model with \CPT conserving generic new physics effects. However, Ref. \cite{kundu3} has taken \CPT violation into account for describing the mode $B^0_s\to J/\psi\,\phi$ and Ref. \cite{kundu4} has discussed about triple products and angular observables for $B\to V_1 V_2$ decays in light of \CPT violation. In this paper, we have revisited the prospect of searching \CPT violation in mixing through $P^0\to V_1 V_2$ decays using helicity-based analysis for  time-dependent differential decay rate. We would also like to emphasize that we have taken a model-independent approach in a sense that we do not specify any definite model that might lead to \CPT violation.

The usual technique to deal with the oscillations of neutral pseudoscalar mesons is to consider a final state $f$ to which both $P^0$ and $\bar P^0$ can decay. If $f$ consists of two vectors, a large number of observables can be constructed from time-dependent differential decay rate depending on the polarization or orbital angular momentum of the final state. But, all of these observables will not be independent to each other and hence there emerge various relations among them. In Ref. \cite{VV2,London:2003rk}, these relations have been discussed in the context of SM scenario only for the modes $B^0_d \text{ or } \bar B^0_d$ decaying to two vectors.  In this paper, we study these relations in presence of \T, \CP and \CPT violations in mixing only. We have confined our analysis to the case where \CPT violation is small compared to the SM amplitude, which is justified based on the data from several experiments \cite{CPTVBmix_BaBar1,CPTVBmix_BaBar2,expts1,expts2,hflav}. Since independent theoretical parameters for this case are more in number than SM scenario, it is expected to obtain lesser number of relations among observables. We find that except four, all the other old relations in SM get violated and new relations appear if \T and \CPT violations in mixing are present. These new relations will hold true even if the \T, \CP and \CPT violations become zero; however, they will not form the complete set of relations in that case as they are less in number. These new relations will break down only if \T, \CP and \CPT violating effects are present in decay too (i.e. direct violation).

The paper is organized as follows. In the next section,
we briefly describe the theoretical formalism for \CPT violation in $P^0-\bar P^0$ mixing and express the time dependent differential decay rate of $P^0$ and $\bar P^0$ in terms of the mixing parameters. In Sec. \ref{Observables}, we construct helicity-dependent observables from the differential decay rates and express them in terms of \T, \CP and \CPT violating parameters assuming \T and \CPT violations in mixing to be very small. We also solve for all the unknown theoretical parameters as functions of the observables. In. Sec. \ref{ObservableRelation}, we establish the independent relations among these observables in SM case and the scenario with the presence of \T and \CPT violations in mixing separately. We also discuss how these relations can help us in distinguishing three different scenarios a) SM case, b) \T, \CP and \CPT violation in mixing and c)  direct violation of \T, \CP and \CPT. Finally, we summarize and conclude in Sec. \ref{Conclusion}.

\section{Theoretical Formalism:}
\label{sec:Theory}

We begin by reviewing the most general formalism for $P^0-\bar P^0 $mixing, in which \CPT and  \T violation are incorporated. This formalism has already been discussed in Ref. \cite{CPTps}; however, for the sake of completeness we present it in this section. In the $(P^0-\bar P^0)$ basis, the generic mixing Hamiltonian can be expressed in terms of two $2\times2$ Hermitian matrices $\mathbf M$ and $\mathbf \Gamma$, respectively the mass and decay matrices, as $\mathbf {M}-(i/2) \mathbf \Gamma$. It should be noticed that the mixing matrix $\mathbf {M}-(i/2) \mathbf \Gamma$ is non-Hermitian and it is justified as the probability of finding $P^0$ and $\bar P^0$ decreases with time due to presence of the non-null decay matrix $\mathbf \Gamma$. Now, since any $2\times2$ matrix can be expanded in terms of three Pauli matrices $\bm\sigma_j$ and identity matrix $\mathbf I$ with complex coefficients, we can write:
\begin{equation}
\label{hamiltonian}
\mathbf M-\frac{i}{2}\mathbf\Gamma=E\sin\theta\cos\phi\,\bm\sigma_1+E\sin\theta\sin\phi\,\bm\sigma_2+E\cos\theta\,\bm\sigma_3-iD\,\mathbf I
\end{equation}
where, $E,\theta,\phi$ and $D$ are complex entities in general. Comparing both sides of this equation, we obtain:
\begin{align}
\label{EDdefs}
&D = \frac{i}{2} (M_{11} + M_{22}) + \frac{1}{4} (\Gamma_{11} + \Gamma_{22}) ~, \nn\\
&E\cos\theta = \frac{1}{2} \, (M_{11} - M_{22}) - \frac{i}{4} (\Gamma_{11} - \Gamma_{22}) ~, \nn\\
&E\sin\theta\cos\phi = {\rm Re} \, M_{12}-\frac{i}{2}{\rm Re} \, \Gamma_{12} ~, \nn\\
&E\sin\theta\sin\phi = -{\rm Im} \, M_{12} + \frac{i}{2} {\rm Im} \, \Gamma_{12} ~.
\end{align}
where $M_{ij}$ and $\Gamma_{ij}$ are $(i,j)$-th elements of $\mathbf M$ and $\mathbf \Gamma$ matrices respectively.

The eigenvectors of the mixing Hamiltonian $\mathbf {M}-(i/2) \mathbf \Gamma$ are the mass eigenstates ($\ket{P_L}$ and $\ket{P_H}$) and they can be expressed as linear combinations of the flavour eigenstates ($\ket{P^0}$ and $\ket{\oln{P^0}}$) as follows:
\begin{eqnarray}
	\label{MState}
	&\ket{P_L}= p_1 \ket{P^0}+q_1 \ket{\bar{P}^0}, \nonumber \\
	&\ket{P_H}= p_2 \ket{P^0}-q_2 \ket{\bar{P}^0},
\end{eqnarray}
where $p_1=N_1\cos{\frac{\theta}{2}},\ q_1=N_1\,e^{i\phi}\sin{\frac{\theta}{2}},\ p_2=N_2\sin{\frac{\theta}{2}}$,  $q_2=N_2\,e^{i\phi}\cos{\frac{\theta}{2}}$ with $N_1, N_2$ being two normalization factors and the $L$,$H$ tags indicate light and heavy physical states, respectively. Since, the physical states, as given by Eq.~\eqref{MState}, depend only on the  parameters $\theta$ and $\phi$, they are called the mixing parameters for \PPbr system. It should be noticed that the physical states are not orthogonal in general since the mixing matrix is non-Hermitian. 

The time evolution of flavour states  ($\ket{B^0} \equiv \ket{B^0(t=0)}$ and $\ket{{\bar B}^0} \equiv \ket{{\bar	B}^0(t=0)}$) is given by:
\begin{align}
	\label{TState}
	&\ket{P^0(t)}=  h_{+} \ket{P^0} + h_{-}\cos{\theta}\ket{P^0}  + h_{-} e^{i \phi} \sin{\theta} \ket{\oln{P^0}}, \nonumber \\
	&\ket{\oln{P^0}(t)}= h_{+} \ket{\oln{P^0}} - h_{-}\cos{\theta}\ket{\oln{P^0}}  + h_{-} e^{-i \phi} \sin{\theta} \ket{P^0},
\end{align}
\vspace*{-10mm}
\begin{align}
\text{\hspace*{-3cm}where, }\qquad  h_+&= e^{-i(M-i \frac{\Gamma}{2})t}  \cos\Big[\Big(\Delta{M}-i\frac{\Delta\Gamma }{2}\Big)\frac{t}{2}\Big], \nonumber\\
	 	h_-&= e^{-i(M-i \frac{\Gamma}{2})t}\,i \sin\Big[\Big(\Delta{M}-i\frac{\Delta\Gamma }{2}\Big)\frac{t}{2}\Big].
\end{align}
Here $M=(M_H + M_L)/2$, $\Delta M=M_H-M_L$, $\Gamma=(\Gamma_H + \Gamma_L)/2$ and $\Delta\Gamma=\Gamma_H-\Gamma_L$ with $M_{L,H}$ and $\Gamma_{L,H}$ to be masses and decay widths of the light and heavy mass eigenstates respectively. 

Let us now consider a final state $f$ to which both $P^0$ and $\bar P^0$ can decay. Using Eq.~\eqref{TState}, the time dependent decay amplitudes for the neutral mesons are given by:
\begin{align}
	\label{TAmplitude}
	\mathcal{A}mp(P^0(t)\rightarrow f)=& h_{+} \mathcal{A}_{f} + h_{-} \cos{\theta} \mathcal{A}_{f} + h_{-} e^{i \phi} \sin{\theta} \oln{\mathcal{A}_{f}},\nonumber \\
	\mathcal{A}mp(\oln{P^0}(t)\rightarrow f)=& h_{+} \oln{\mathcal{A}_{f}} - h_{-} \cos{\theta} \oln{\mathcal{A}_{f}} + h_{-} e^{-i \phi} \sin{\theta} \mathcal{A}_{f}, 
\end{align} 
where $\mathcal{A}_f= \bra{f}\mathcal{H}_{\Delta F=1}\ket{P^0} \text{ and }\oln{\mathcal{A}_{f}}=\bra{f}\mathcal{H}_{\Delta F=1}\ket{\oln{P^0}}$.  Hence, the decay rates $\Gamma_f(P^0(t)\rightarrow f)$ and $\oln{\Gamma}_f(\oln{P^0}(t)\rightarrow f)$ can be expressed as:
\small
\bea
	\label{TGamma}
	\frac{d\Gamma}{dt}(P^0(t)\to   f) &=& \frac12 e^{-\Gamma t} 
	\left[
	\sinh\left(\Delta\Gamma t/2\right)  \left\{ 2\text{Re} 
	\left( \cos\theta |\mathcal{A}_{f}|^2 +e^{i \phi} \sin\theta 
	\mathcal{A}_{f}^{*} \bar{\mathcal{A}}_{f} \right) \right\}
	\right. \nn \\
	&& \hskip-1truein  +~\cosh\left(\Delta\Gamma t/2\right) \left\{ |\mathcal{A}_f |^2 + |\cos\theta |^2|\mathcal{A}_f |^2 + |e^{i \phi}\sin\theta |^2|\bar{\mathcal{A}}_{f} |^2 
	+ 2 \text{Re}\left(e^{i \phi} \cos\theta^{*} \sin\theta 
	\mathcal{A}_{f}^{*} \bar{\mathcal{A}}_{f} \right) \right\} \nn\\
	&& \hskip-1truein  +~\cos(\Delta M t) \left\{ |\mathcal{A}_f |^2 - |\cos\theta |^2|\mathcal{A}_f |^2 - |e^{i \phi}\sin\theta |^2|\bar{\mathcal{A}}_{f} |^2 - 2 \text{Re}\left(e^{i \phi} \cos\theta^{*} \sin\theta 
	\mathcal{A}_{f}^{*} \bar{\mathcal{A}}_{f} \right)
	\right\} \nn\\
	&& \left. -~\sin(\Delta Mt) \left\{ 2 \text{Im} \left( \cos\theta |\mathcal{A}_{f}|^2 + e^{i \phi} \sin\theta \mathcal{A}_{f}^{*}\bar{\mathcal{A}}_{f}
	\right) \right\}
	\right] ~,
	\eea
	\bea
	\label{TGammaPar}
	\frac{d\Gamma}{dt}({\bar P}^0(t)\to  f) &=& \frac12 e^{-\Gamma t} \left[
	\sinh\left(\Delta\Gamma t/2\right) \left\{ 2\text{Re} 
	\left( -\cos\theta^{*} |\bar{\mathcal{A}}_{f}|^2 + e^{i \phi^*} 
	\sin\theta^{*} \mathcal{A}_{f}^{*} \bar{\mathcal{A}}_{f} \right) \right\} \right. \nn \\
	&& \hskip-1truein +~\cosh\left(\Delta\Gamma t/2\right) \left\{ |\oln{\mathcal{A}_f} |^2 + |\cos\theta |^2|\oln{\mathcal{A}_f} |^2 + | e^{-i \phi}\sin\theta |^2|\mathcal{A}_{f} |^2 
	- 2 \text{Re}\left( e^{i \phi^*} \cos\theta \sin\theta^{*} \mathcal{A}_{f}^{*} \bar{\mathcal{A}}_{f} \right)
	\right\} \nn\\
	&& \hskip-1truein +~\cos(\Delta M t) \left\{ |\oln{\mathcal{A}_f} |^2 - |\cos\theta |^2|\oln{\mathcal{A}_f} |^2 - | e^{-i \phi}\sin\theta |^2|\mathcal{A}_{f} |^2 + 2 \text{Re}\left( e^{i \phi^*} \cos\theta \sin\theta^{*} 
	\mathcal{A}_{f}^{*} \bar{\mathcal{A}}_{f} \right)
	\right\} \nn\\
	&& \left. +~\sin(\Delta Mt) \left\{ 2 \text{Im} \left( -\cos\theta^{*}  
	|\bar{\mathcal{A}}_{f}|^2 +  e^{i \phi^*} \sin\theta^{*} 
	\mathcal{A}_{f}^{*} \bar{\mathcal{A}}_{f}
	\right) \right\} \right] ~.
\eea
\normalsize

\section{Observables:}
\label{Observables}
\subsection{Decay Rates:}
\label{DecayRates}

Any final state consisting of two vectors can have three different values for orbital angular momentum quantum number $\{0,1,2\}$ which correspond to the polarization states $\{0,\perp,\parallel\}$, respectively. As we are not considering \CPT violation in decay, we can express the decay amplitudes for modes and conjugate modes in terms of transversity amplitudes as \cite{VV1,VV2,London:2003rk,kundu4}:
\begin{align}
\label{VVAmplitude}
\mathcal{A}_{f}(P^0\rightarrow V_{1}V_{2}) =& \mathcal{A}_{0} g_{0} + \mathcal{A}_{\pl} g_{\pl} + i\mathcal{A}_{\perp} g_{\perp}= \sum_{\lambda}\mathcal{A}_{\lambda}g_{\lambda}\zeta_{\lambda} \nonumber~, \\
\widebar{\mathcal{A}_{f}}(\oln{P^0}\rightarrow V_{1}V_{2}) =& \widebar{\mathcal{A}_{0}} g_{0} + \widebar{\mathcal{A}_{\pl}} g_{\pl} - i\widebar{\mathcal{A}_{\perp}} g_{\perp}= \sum_{\lambda}\widebar{\mathcal{A}_{\lambda}}g_{\lambda}\zeta_{\lambda}^{*}~.                           
\end{align}
where the helicity index $\lambda$ takes the value $\{0,\parallel,\perp\}$ and $\zeta_{\lambda}$ takes the value $\{1,1,i\}$ for these three helicities respectively. The factor $g_\lambda$  are the coefficients of helicity amplitudes ($\A_\lambda$ or $\oln{\A_\lambda}$) in linear polarization basis and only depend on kinematic angles~\cite{Sinha:1997zu}. In absence of direct violation for \CP, \T and \CPT, these helicity amplitudes can be expressed as:
\begin{equation}
\label{AAbar}
\mathcal{A}_{\lambda}=\widebar{\mathcal{A}_{\lambda}}=a_\lambda e^{i\delta_{\lambda}}.
\end{equation}
where, $a_\lambda$ and $\delta_\lambda$ are two real quantities indicating the magnitudes and phases for different helicity amplitudes. 

Now, using Eq.~\eqref{TGamma}-\eqref{AAbar}, the time-dependent decay rates for $P^0\rightarrow V_1V_2$ and $\bar{P}^0\rightarrow V_1V_2$ modes can be written as~\cite{VV1,VV2,London:2003rk, kundu3, kundu4}:
\small
\begin{align}
\label{VVDecayRate}
\begin{split}
&\frac{d\Gamma}{dt}(P^0(t)\to V_1V_2\big) = \\
&e^{-\Gamma t}\sum_{\lambda\leq\sigma}\Bigg[\Lambda_{\lambda\sigma}\cosh\Big(\frac{\Delta\Gamma t}{2}\Big)+\eta_{\lambda\sigma}^{}\sinh\Big(\frac{\Delta\Gamma t}{2}\Big)+\Sigma_{\lambda\sigma}\cos\big(\Delta M t\big)-\rho_{\lambda\sigma}^{}\sin\big(\Delta M t\big)\Bigg]g_\lambda g_\sigma~,
\end{split} \\
\label{ConjVVDecay}
\begin{split}
&\frac{d\Gamma}{dt}(\oln{P}^0(t)\to V_1V_2\big) = \\
&e^{-\Gamma t}\sum_{\lambda\leq\sigma}\Bigg[\oln{\Lambda}_{\lambda\sigma}\cosh\Big(\frac{\Delta\Gamma t}{2}\Big)+\oln{\eta}_{\lambda\sigma}^{}\sinh\Big(\frac{\Delta\Gamma t}{2}\Big)+\oln{\Sigma}_{\lambda\sigma}\cos\big(\Delta M t\big)+\oln{\rho}_{\lambda\sigma}^{}\sin\big(\Delta M t\big)\Bigg]g_\lambda g_\sigma~.
\end{split}
\end{align}
\normalsize
where both $\lambda$ and $\sigma$ take the value $\{0,\parallel,\perp\}$.

From Eq.~\eqref{VVDecayRate} we see that for each of the helicity combination, there are four observables $(\Lambda_{\ls},\eta_{\ls},\Sigma_{\ls}, \rho_{\ls})$ and six such helicity combinations are possible. Hence, we get total 24 observables for $P^0\rightarrow V_1V_2$ mode. Similarly, there will be 24 different observables $(\bar{\Lambda}_{\ls},\bar{\eta}_{\ls},\bar{\Sigma}_{\ls}, \bar{\rho}_{\ls})$ for $\bar{P}^0\rightarrow V_1V_2$ mode too. These observables can be measured by performing a time dependent angular analysis of $P^0(t)\rightarrow V_1V_2$ and $\bar P^0(t)\rightarrow V_1V_2$ \cite{VV1,VV2,London:2003rk}. The procedure described in Ref. \cite{kundu4} can be helpful in this regard.  On the other hand, probing polarizations of the final state particles may also aid in measurement of these observables. One important point to notice here is that Ref.~\cite{VV1,VV2,London:2003rk} did not consider  $\sinh\Big(\frac{\Delta\Gamma t}{2}\Big)$ terms in the decays of $B^0_d$ and $\bar B^0_d$ since $\Delta\Gamma$ is consistent with zero \cite{pdg} for these modes.
In that case, $\eta_{\lambda\sigma}$ and $\bar\eta_{\lambda\sigma}$ remain undetermined and one should work with remaining $(18+18)=36$ observables for a mode and its conjugate mode. However, since we are considering a general scenario here, we keep all the terms and proceed.  

\subsection{Parametric Expansion:}
\label{ParametricExpansion}

In Ref.~\cite{lee} (pgs. 349$-$358), T.D. Lee discusses the \CPT and \T properties of $\mathbf M$ and $\mathbf \Gamma$ matrices. First, if \CPT invariance holds, then, independently of \T symmetry,
\begin{equation}
\label{CPTgood}
M_{11} = M_{22} ~,~~ \Gamma_{11} = \Gamma_{22} \quad \Longrightarrow \theta = \frac{\pi}{2} \quad \text{(Using Eq. \eqref{EDdefs})}.
\end{equation}
In addition, if \T invariance holds, then, independently of \CPT symmetry,
\begin{equation}
\label{Tgood}
\frac{\Gamma_{12}^*}{\Gamma_{12}} = \frac{M_{12}^*}{M_{12}}
 \Longrightarrow {\rm Im} \, \phi = 0 ~\quad \text{(Using Eq. \eqref{EDdefs})}.
\end{equation}
Hence, incorporating \T, \CP and \CPT violation in  \PPbr mixing, we parametrize $\theta$ and $\phi$ as \cite{CPTps}:
\begin{equation}
\label{epsdef}
\theta=\frac{\pi}{2}+\epsilon_1+i\epsilon_2\; \text{ and }\; \phi=-2\beta+i\epsilon_3
\end{equation}
where, $\beta$ is the \CP violating weak phase, $\epsilon_1\text{ and }\epsilon_2$ are \CPT violating parameters and $\epsilon_3$ is \T violating parameter. The notation of Belle, BaBar and LHCb  collaborations \cite{CPTVBmix_BaBar1,CPTVBmix_BaBar2,expts1,expts2} is a bit different from ours; however, the two notations are related to each other by the following transformation \cite{CPTps}:
\beq
\cos\theta \leftrightarrow  -z ~,~~ \sin\theta \leftrightarrow \sqrt{1-z^2} ~,~~ e^{i\phi} \leftrightarrow \frac{q}{p} ~,
\eeq
\beq
\label{epsdefs}
\text{or, equivalently: }\quad \e_1 = {\rm Re}(z) ~,~~ \e_2 = {\rm Im}(z) ~,~~ \e_3 = 1-\Big|\frac{q}{p}\Big| ~. 
\eeq


Now, Comparing Eq.~\eqref{TGamma} to Eq.~\eqref{VVDecayRate} one can easily infer that all of the observables will be functions of the complex quantities $\theta$ and $\phi$. As \T and \CPT violations are expected to be very small \cite{CPTVBmix_BaBar1,CPTVBmix_BaBar2,expts1,expts2,hflav}, we can expand all the observables in terms of $\epsilon_j$ ($j\in\{1,2,3\}$). So, using Eq.\eqref{AAbar} and Eq.~\eqref{epsdef}, we expand all of the 24 helicity dependent observables for $P^0\to V_1 V_2$ in terms of $\e_j$ ($j\in\{1,2,3\}$) keeping up to the linear terms as following:
\begin{align}
	\label{Lambda}
	&\Lambda_{ii} = a_i^2 \big( 1 -\e_3 - \e_1\c2b + \e_2 \s2b \big), \nonumber \\
	&\Lambda_{\perp\perp} = a_\perp^2 \big( 1- \e_3 + \e_1\c2b - \e_2\s2b  \big), \nonumber \\
	&\Lambda_{0\pl} = 2 a_0 a_\pl \cos(\Delta_0 - \Delta_\pl) \big( 1- \e_3 - \e_1\c2b + \e_2\s2b \big), \nonumber \\
	&\Lambda_{\perp i} = 2 a_\perp a_i \big( (\e_2\c2b + \e_1\s2b)\cos\Delta_i + \e_3\sin\Delta_i \big), \\
	\nonumber\\
	\label{eta}
	&\eta_{ii} = -a_i^2 \big( \e_1 - \c2b +\e_3\c2b \big), \nonumber  \\
	&\eta_{\perp\perp} = -a_\perp^2 \big( \e_1 + \c2b - \e_3\c2b \big), \nonumber \\
	&\eta_{0\pl} = -2 a_0 a_\pl \cos(\Delta_0 - \Delta_\pl) \big( \e_1 - \c2b + \e_3 \c2b \big), \nonumber \\
	&\eta_{\perp i} = -2 a_\perp a_i \big( (1-\e_3)\s2b\cos\Delta_i + \e_1\sin\Delta_i \big), \\
	\nonumber\\
	\label{Sigma}
	&\Sigma_{ii} = a_i^2 \big( \e_3 + \e_1\c2b - \e_2\s2b \big), \nonumber \\
	&\Sigma_{\perp\perp} = a_\perp^2 \big( \e_3 - \e_1\c2b + \e_2\s2b \big), \nonumber \\
	&\Sigma_{0\pl} = 2 a_0 a_\pl \cos(\Delta_0-\Delta_\pl) \big( \e_3 + \e_1\c2b - \e_2\s2b \big),  \nonumber \\
	&\Sigma_{\perp i} = -2 a_\perp a_i \big( (\e_2\c2b + \e_1\s2b)\cos\Delta_i -(1 - \e_3)\sin\Delta_i \big), \\
	\nonumber\\
	\label{rho}
	&\rho_{ii} = - a_i^2 \big( \e_2 + \s2b - \e_3\s2b \big), \nonumber \\
	&\rho_{\perp\perp} = - a_\perp^2 \big( \e_2 - \s2b + \e_3\s2b \big), \nonumber \\
	&\rho_{0\pl} = -2 a_0 a_\pl \cos(\Delta_0 - \Delta_\pl) \big( \e_2 + \s2b - \e_3\s2b \big), \nonumber \\
	&\rho_{\perp i} = -2 a_\perp a_i \big( (1-\e_3)\c2b\cos\Delta_i + \e_2\sin\Delta_i \big) ,
\end{align}
where $i \in \{0, \pl\}$ and $\Delta_i = \delta_i - \delta_\perp$. Similarly, it's also possible to expand the observables of the conjugate mode $\bar{P}^0\to V_1V_2$ in terms of $\epsilon_j$ (given in Appendix \ref{App}).

\subsection{Solutions:}
\vspace{5mm}

As can be seen from the expansion of observables, given by Eq.~\eqref{Lambda}-\eqref{rho}, there  are total  9 unknown parameters (i.e. 3 of $a_\lambda$, 3 of $\epsilon_j$, 2 of $\Delta_i$ and $\beta$). In SM case, there are six unknown parameters (3 of $a_\lambda$, 2 of $\Delta_i$ and $\beta$), as stated in Ref. \cite{VV2,London:2003rk}; however, for our scenario, we have three extra parameters emerging due to \T and \CPT violation in mixing namely $\epsilon_{1,2,3}$, thus resulting in nine theoretical parameters. It should be noted that Ref. \cite{VV2,London:2003rk} originally deal with SM scenario plus \CP violation in decay, not \T and \CPT violation in mixing; hence, in addition to six unknown SM parameters, they have three more amplitudes ($b_\lambda$), three more strong phases ($\delta_\lambda^b$) and one extra weak phase related to the \CP violating part of the decay amplitudes ($\mathcal{A}_\lambda$ or $\bar{\mathcal{A}}_\lambda$). Now, we go back to our scenario and solve the nine theoretical parameters in terms of observables as follows:
\begin{align}
	\label{alambdaSol}
	&a_\lambda^{}=\sqrt{\Lambda_{\lambda\lambda}+\Sigma_{\lambda\lambda}}, \\
	\label{eps1Sol}
	& \e_1 = -\frac{1}{2}\Big(\frac{\eta_{ ii}^{}}{\Lambda_{ ii}+\Sigma_{ ii}}+\frac{\eta_{\perp\perp}^{}}{\Lambda_{\perp\perp}+\Sigma_{\perp\perp}}\Big), \\
	\label{eps2Sol}
	& \e_2 = -\frac{1}{2}\Big(\frac{\rho_{ ii}^{}}{\Lambda_{ ii}+\Sigma_{ ii}}+\frac{\rho_{\perp\perp}^{}}{\Lambda_{\perp\perp}+\Sigma_{\perp\perp}}\Big), \\
	\label{eps3Sol}
	& \e_3 = \frac{1}{2}\Big(\frac{\Sigma_{ ii}^{}}{\Lambda_{ ii}+\Sigma_{ ii}}+\frac{\Sigma_{\perp\perp}^{}}{\Lambda_{\perp\perp}+\Sigma_{\perp\perp}}\Big), \\
	\label{SinPetaSol}
	&\sin2\beta=
	-\frac{1}{2}\Big(\frac{\rho_{ii}}{\Lambda_{ii}}-\frac{\rho_{\perp\perp}}{\Lambda_{\perp\perp}}\Big),\\
	\label{CosPetaSol}
	&\cos2\beta=
	\frac{1}{2}\Big(\frac{\eta_{ii}^{}}{\Lambda_{ii}}-\frac{\eta^{}_{\perp\perp}}{\Lambda_{\perp\perp}}\Big),\\
	\label{CosDeltaDeltaSol}
	&\cos(\Delta_0-\Delta_\parallel)=\frac{1}{2}\Big[\frac{\Lambda_{0\parallel}+\Sigma_{0\parallel}}{\sqrt{\Lambda_{00}+\Sigma_{00}}\sqrt{\Lambda_{\parallel\parallel}+\Sigma_{\parallel\parallel}}}\Big],\\
	\label{SinDeltaSol}
	& \sin\Delta_i = \frac{1}{2}\Big[\frac{\Lambda_{\perp i}+\Sigma_{\perp i}}{\sqrt{\Lambda_{ii}+\Sigma_{ii}}\sqrt{\Lambda_{\perp\perp}+\Sigma_{\perp\perp}}}\Big], \\
	\label{CosDeltaSol}
	&\cos\Delta_i=X_i\, \Lambda_{ ii}\,\Lambda_{\perp\perp}\Big[\frac{\sqrt{\Lambda_{ii}+\Sigma_{ii}}\sqrt{\Lambda_{\perp\perp}+\Sigma_{\perp\perp}}}{\Lambda_{\perp\perp}\Sigma_{ ii}+\Lambda_{ii}\Sigma_{\perp\perp}+2\Lambda_{\perp\perp}\Lambda_{ii}}\Big] ,
\end{align}
\begin{align}
\hspace*{-0.3cm}\text{where, }{\label{Xi}
	X_i=\Big[\frac{\big(\Lambda_{\perp i}-\Sigma_{\perp i}\big)\big(\Lambda_{\perp\perp}\Sigma_{ ii}+\Lambda_{ ii}\Sigma_{\perp\perp}\big)+2\big(\Lambda_{ ii}\Lambda_{\perp\perp}\Lambda_{\perp i}-\Sigma_{ ii}\Sigma_{\perp\perp}\Sigma_{\perp i}\big)}{\big(\eta_{\perp\perp}^{} \rho_{ii}^{}-\eta_{ii}^{} \rho_{\perp\perp}^{}\big)\big(\Lambda_{ii}+\Sigma_{ii}\big)\big(\Lambda_{\perp\perp}+\Sigma_{\perp\perp}\big)}\Big]}, 
\end{align}
with $\lambda\in\{0,\pl,\perp\}$ and $i \in \{0, \pl\}$. In principle, we should present only 9 equations as the solutions for 9 unknown parameters. But, we have listed more than 9 relations from Eq.~\eqref{alambdaSol} to Eq.~\eqref{CosDeltaSol} because the observables involve several angular parameters. Actually, to specify any angular variable without any ambiguity, one must quantify both $\sin$ and $\cos$ of that angle. However, as can be seen in section \ref{T-CPT},  the extra equations will result in some relations among observables by applying various trigonometric identities.

\section{Observable Relations:}
\label{ObservableRelation}
\subsection{SM relations:}
\vspace{5mm}
 In SM scenario, all of the three $\epsilon_j$ become zero and  there remain only 6 unknown parameters (3 of $a_\lambda$, 2 of $\Delta_i$ and $\beta$) in the theory. But the number of observables for $P^0\to V_1V_2$ mode is 24. Hence,  18 independent relations among observables must emerge and they are the following:
\begin{align}
\label{sig-lam}
&\Sigma_{\lambda\lambda}=0, \text{ } \Sigma_{0\parallel}=
0,\text{ } \Lambda_{\perp i}=0\,, \\
\label{rhobylam}
&\frac{\rho_{ii}^{}}{\Lambda_{ii}}= \frac{\rho_{0\parallel}^{}}{\Lambda_{0\parallel}}=-\frac{\rho_{\perp\perp}^{}}{\Lambda_{\perp\perp}}\,,\\
\label{rhopisq}
& \frac{\rho_{\perp i}^2}{4\Lambda_{\perp\perp}\Lambda_{ii}-\Sigma_{\perp i}^2}=\frac{\Lambda_{0\parallel}^2-\rho_{0\parallel}^2}{\Lambda_{0\parallel}^2}\,,\\
\label{lam0par}
&\Lambda_{0\parallel}=\frac{1}{2\Lambda_{\perp\perp}}\Big[\Sigma_{\perp 0}\Sigma_{\perp\parallel}+\rho_{\perp 0}^{}\rho_{\perp\parallel}^{}\Big(\frac{\Lambda_{0\parallel}^2}{\Lambda_{0\parallel}^2-\rho_{0\parallel}^2}\Big)\Big]\,.\\
\label{etabylam}
&\frac{\eta_{ii}^{}}{\Lambda_{ii}}= \frac{\eta_{0\parallel}^{}}{\Lambda_{0\parallel}}=-\frac{\eta_{\perp\perp}^{}}{\Lambda_{\perp\perp}}\,,\\
\label{etabyrho}
&\frac{\eta_{\perp i}^{}}{\rho_{\perp i}^{}}+\frac{\eta_{0\parallel}^{}}{\rho_{0\parallel}^{}}=0\,,\\
\label{lamsq}
& \eta_{0\parallel}^2+\rho_{0\parallel}^2=\Lambda_{0\parallel}^2\, ,
\end{align}
with $\lambda\in\{0,\pl,\perp\}$ and $i \in \{0, \pl\}$. Here, Eq.~\eqref{sig-lam} contains 6 relations (for 3 different $\lambda$ and two different $i$), Eq.~\eqref{rhobylam} and  Eq.~\eqref{etabylam} contain 3 relations each (for two different $i$) whereas there are two relations ( for two different $i$) inside each of Eq.~\eqref{rhopisq} and Eq.~\eqref{etabyrho}. 

However, for vanishing $\Delta\Gamma$, only 18 observables will be accessible to us (as discussed in the section \ref{DecayRates}) and hence, in that case, we should obtain 12 independent relations among observables. Those 12 relations are given by Eq.~\eqref{sig-lam} $-$ Eq. \eqref{lam0par}, as discussed in Ref. \cite{VV2,London:2003rk}.

 One important point to state is that one can use the solutions, given by Eq.~\eqref{alambdaSol}-\eqref{SinDeltaSol}, in the SM scenario also. But, $X_i$, given by Eq.~\eqref{Xi}, takes the form $\frac{0}{0}$ in this case and it causes problem in finding $\cos\Delta_i$ from Eq.~\eqref{CosDeltaSol}. Still, one can express $\cos\Delta_i$ ($i \in \{0, \pl\}$) in this scenario as following:
 \begin{equation}
 \label{CosDelta3}
 \cos\Delta_i=-\Big(\frac{\Lambda_{0\parallel}^{}\rho_{\perp i}^{}}{2\eta_{0\parallel}\sqrt{\Lambda_{ii}\Lambda_{\perp\perp}}}\Big)\,,
 \end{equation}
which can easily be verified by substituting vanishing $\epsilon_j$ into the Eq.~\eqref{Lambda} $-$ Eq.~\eqref{rho}. Hence, using Eq. \eqref{CosDeltaSol}, Eq. \eqref{sig-lam} and Eq. \eqref{CosDelta3}, one can write $X_i$ ($i \in \{0, \pl\}$) in the limit $\epsilon_j\to 0$ ($j\in\{1,2,3\}$) as:
\begin{equation}
 X_i=-\Big(\frac{\Lambda_{0\parallel}\rho_{\perp i}}{\eta_{0\parallel}\Lambda_{ii}\Lambda_{\perp\perp}}\Big)\,.
\end{equation}

Nevertheless, we shall see in the next section that most of these 18 relations from Eq. \eqref{sig-lam} $-$ Eq. \eqref{lamsq} will get violated if \T and \CPT violations in mixing are also  present. On the other hand, if there exists direct violation of \T, \CP or \CPT instead of \T and \CPT violating effects in mixing,  then also most of these relations get violated. Hence, it is impossible to infer from this set of relations whether \CPT violation (if it exists at all) is present in mixing or in decay.

\subsection{\T and \CPT violation:}
\label{T-CPT}
\vspace*{5mm}
In addition to the \CP violating weak phase if there exist \T and \CPT violation in mixing, we have 9 unknown theoretical parameters (3 of $\epsilon_j$, 3 of $a_\lambda$, 2 of $\Delta_i$ and $\beta$). But the number of observables is still 24. So, there should appear $(24-9)=15$ number of relations among observables. In order to find them, we substitute the solutions of unknown parameters, given by Eq.~\eqref{alambdaSol}$-$\eqref{CosDeltaSol}, back to the expansion of observables, given by Eq.~\eqref{Lambda}-\eqref{rho}. Thus we get 11 independent relations, which are given below:
\begin{align}
\label{lam-sig-rho-eta}
	&\frac{\Lambda_{0\parallel}}{\Lambda_{ii}}=\frac{\Sigma_{0\parallel}}{\Sigma_{ii}}=\frac{\rho_{0\parallel}^{}}{\rho_{ii}^{}}=\frac{\eta_{0\parallel}^{}}{\eta_{ii}^{}},\\
	\label{rho-eta-lam}
	&\frac{\rho_{0\parallel}^2+\eta_{0\parallel}^2}{\Lambda_{0\parallel}^2}=\frac{\rho_{ \perp\perp}^2+\eta_{ \perp\perp}^2}{\Lambda_{ \perp\perp}^2} ,
\end{align}
\vspace*{-7mm}\small
\begin{align}
    \label{etapi}
	&\eta_{\perp i}=\frac{1}{2}\Bigg[\frac{\Sigma_{\perp i}}{\Lambda_{ii}\Lambda_{\perp\perp}}\Big\{\eta_{\perp\perp}^{}\big(\Lambda_{ii}+\Sigma_{ii}\big)+\eta_{ii}^{}\big(\Lambda_{\perp\perp}+\Sigma_{\perp\perp}\big)\Big\}+X_i\Big\{\Lambda_{\perp\perp}\rho_{ii}^{}-\Lambda_{ii}\rho_{\perp\perp}^{}\Big\}\Bigg], \\
	\label{rhopi}
	&\rho_{\perp i}=\frac{1}{2}\Bigg[\frac{\Sigma_{\perp i}}{\Lambda_{ii}\Lambda_{\perp\perp}}\Big\{\rho_{\perp\perp}^{}\big(\Lambda_{ii}+\Sigma_{ii}\big)+\rho_{ii}^{}\big(\Lambda_{\perp\perp}+\Sigma_{\perp\perp}\big)\Big\}-X_i\Big\{\Lambda_{\perp\perp}\eta_{ii}^{}-\Lambda_{ii}\eta_{\perp\perp}^{}\Big\}\Bigg],
\end{align}
\normalsize
with $i \in \{0, \pl\}$. It should be noticed that there are 6 independent relations in Eq.~\eqref{lam-sig-rho-eta}, 2 relations in Eq.~\eqref{etapi} and 2 relations in Eq.~\eqref{rhopi}.

There are four more such independent relations among observables which emerges due to the following trigonometric identities:
\begin{align}
\label{sinsq+cossq}
	\sin^2 \alpha+\cos^2 \alpha&=1\quad \text{(where $\alpha=\Delta_0$, $\Delta_\parallel$ or $2\beta)$},\\
	\label{cos0-para}
	\cos(\Delta_0-\Delta_\parallel)&=\cos\Delta_0 \cos\Delta_\parallel+\sin\Delta_0 \sin\Delta_\parallel.
\end{align}
Substituting expressions for different angular variables from Eq.~\eqref{SinPetaSol}$-$Eq.~\eqref{CosDeltaSol} into the above trigonometric identities, given by Eq.~\eqref{sinsq+cossq} and Eq.~\eqref{cos0-para}, we get the remaining four relations as:
\small
\begin{align}
\label{SinsqCossqDelta}
\Big[\frac{(\Lambda_{\perp i}+\Sigma_{\perp i})^2}{(\Lambda_{ii}+\Sigma_{ii})(\Lambda_{\perp\perp}+\Sigma_{\perp\perp})}\Big]+
4 X_i^2\, \Lambda_{ ii}^2\,\Lambda_{\perp\perp}^2\Big[\frac{(\Lambda_{ii}+\Sigma_{ii})(\Lambda_{\perp\perp}+\Sigma_{\perp\perp})}{(\Lambda_{\perp\perp}\Sigma_{ ii}+\Lambda_{ii}\Sigma_{\perp\perp}+2\Lambda_{\perp\perp}\Lambda_{ii})^2}\Big] =4,
\end{align}
\normalsize
\begin{align}
\label{SinsqCossqBeta}
\Big(\frac{\rho_{00}}{\Lambda_{00}}-\frac{\rho_{\perp\perp}}{\Lambda_{\perp\perp}}\Big)^2+
\Big(\frac{\eta_{00}^{}}{\Lambda_{00}}-\frac{\eta^{}_{\perp\perp}}{\Lambda_{\perp\perp}}\Big)^2=4,
\end{align}
\small
\begin{align}
\label{Delta0minuspara}
\Big(\Lambda_{0\parallel}+\Sigma_{0\parallel}&\Big)-\frac{1}{2}\Big[\frac{(\Lambda_{\perp 0}+\Sigma_{\perp 0})(\Lambda_{\perp \parallel}+\Sigma_{\perp \parallel})}{(\Lambda_{\perp\perp}+\Sigma_{\perp\perp})}\Big]\nonumber \\
&=\Big[\frac{2X_0 X_\parallel\, \Lambda_{ 00}\Lambda_{\parallel\parallel}\,\Lambda_{\perp\perp}^2(\Lambda_{00}+\Sigma_{00})(\Lambda_{\parallel\parallel}+\Sigma_{\parallel\parallel})(\Lambda_{\perp\perp}+\Sigma_{\perp\perp})}{(\Lambda_{\perp\perp}\Sigma_{00}+\Lambda_{00}\Sigma_{\perp\perp}+2\Lambda_{\perp\perp}\Lambda_{00})(\Lambda_{\perp\perp}\Sigma_{\parallel\parallel}+\Lambda_{\parallel\parallel}\Sigma_{\perp\perp}+2\Lambda_{\perp\perp}\Lambda_{\parallel\parallel})}\Big],
\end{align}
\normalsize
with $i \in \{0, \pl\}$. The Eq.~\eqref{SinsqCossqDelta} contains two relations (for two different $i$). However, it should be noticed that though $\sin 2\beta$ and $\cos 2\beta$ can be expressed in two ways using the helicities $0$ and $\parallel$ separately (as shown in Eq. \eqref{SinPetaSol} and Eq. \eqref{CosPetaSol}), we obtain only one relation among observables from the trigonometric identity: $\sin^2 2\beta+\cos^2 2\beta=1$. It happens because of the fact that Eq.~\eqref{lam-sig-rho-eta} ensures: $(\rho_{00}/\Lambda_{00})=(\rho_{\parallel\parallel}/{\Lambda_{\parallel\parallel}})$ and $(\eta_{00}/\Lambda_{00})=(\eta_{\parallel\parallel}/{\Lambda_{\parallel\parallel}})$.

However, one should keep in mind that the relations in Eq.~\eqref{lam-sig-rho-eta}-\eqref{cos0-para} will not hold true for all orders in $\epsilon_j$ as we are computing the observables perturbatively up to the first order in $\epsilon_j$. The corrections to these relations are quadratic or of higher order in $\epsilon_j$ and hence can be neglected for sufficiently small values of $\epsilon_j$. Now, if one wants to check the validity of the 18 relations of last section (given by Eq.~\eqref{sig-lam} $-$ Eq.~\eqref{lamsq}) in this scenario, he/she would find $\epsilon_j$ order correction terms in 14 of them. The 4 relations, which remain intact in both the scenarios are: $(\rho_{ii}^{}/\Lambda_{ii})= (\rho_{0\parallel}^{}/\Lambda_{0\parallel})$ and $(\eta_{ii}^{}/\Lambda_{ii})= (\eta_{0\parallel}^{}/\Lambda_{0\parallel})$ which can easily be observed from  Eq.~\eqref{rhobylam}, Eq.~\eqref{etabylam} and Eq. \eqref{lam-sig-rho-eta}.

It should be noted that the 15 relations of this section (Eq. \eqref{lam-sig-rho-eta} $-$ Eq. \eqref{rhopi} and Eq.~\eqref{SinsqCossqDelta} $-$ Eq. \eqref{Delta0minuspara}) hold true even if all $\epsilon_j$ become zero.  It can be verified straightforwardly by setting $\epsilon_j=0$ in parametric expansion of observables (Eq. \eqref{Lambda} $-$ Eq. \eqref{rho}) and then substituting those expressions for observables into these 15 relations. But it does not mean that we have 15 more independent relations in SM case. Because one can easily check that the 18 relations in last section automatically satisfy the 15 relations of this section. In other words, the 18 relations of previous section is embedded in a complicated form inside the 15 relations of present section. However, as discussed in last section, one has to be careful in dealing with $X_i$ while verifying since it takes $\frac{0}{0}$ form in SM scenario. 

Now, if direct violations of \T, \CP and \CPT are present in the decay mode, most of these 15 relations will not hold true and that can be used as a smoking gun signal of confirming those effects. In that case, the 18 relations of SM scenario will be disobeyed too.  On the other hand, if these 15 relations are satisfied, then one becomes sure that there is no direct violations of \T, \CP and \CPT, but it cannot be confirmed whether \T and \CPT violations in mixing are present or not since those 15 relations are satisfied on both the occasions. In this circumstance, the validity of the 18 relations in last section should be examined. If those 18 relations hold true, it would signify the absence of \T and \CPT violation in mixing and if they get violated, the presence of them will be confirmed.

There is another way to confirm the existence of \T, \CP and \CPT violation in decay. In this analysis, we have used the observables of $P^0\to V_1V_2$ mode only for solving all of the 9 unknown parameters, as shown in Eq.~\eqref{alambdaSol} $-$ \eqref{CosDeltaSol}. Similarly, it is also possible to solve them by using the observables of $\bar{P}^0\to V_1V_2$ mode, as given in the Appendix \ref{App}. These two sets of solutions should match numerically in the absence of NP effects in decay. Hence, significant deviations in the numerical values of the  9 unknown parameters from these two sets of solutions will definitely indicate sizeable contributions of \T, \CP and \CPT violations in decay.


\section{Conclusion:}
\label{Conclusion}
In conclusion, we have studied the behaviour of observables for neutral meson decaying to two vectors in the presence of \T, \CP and \CPT violation in mixing. Polarizations of final state with two vectors  provide us a large number of observables in these modes. We choose the final state in such a way that both $P^0$ and $\bar P^0$ can decay to it.
We establish the complete set of 15 relations among observables which must be obeyed if there do not exist any direct violations of \T, \CP and \CPT and these relation can be used as the smoking gun signal to confirm their presence or absence. In addition to that we also listed the full set of 18 relations among observables which should be satisfied if there is no violation of \T and \CPT in mixing of $P^0-\bar P^0$ and these relations can be used to probe their existence unambiguously. 

\section*{Acknowledgement}
The authors thank Rahul Sinha, Nita Sinha and David London for useful discussions. AK thanks SERB India,  grant no: SERB/PHY/F181/2018-19/G210, for support.

\appendix
\section{Appendix}
\label{App}

Expansion of observables for $\bar{P}^0\to V_1V_2$ mode in terms of $\epsilon_j$ ($j\in\{1,2,3\}$) are given by:
\begin{align}
\label{Lambdabar}
&\oln{\Lambda}_{ii} = a_i^2 \big( 1 + \e_3 + \e_1\c2b + \e_2 \s2b \big)~, \nonumber \\
&\oln{\Lambda}_{\perp\perp} = a_\perp^2 \big( 1 + \e_3 - \e_1\c2b - \e_2\s2b  \big)~, \nonumber \\
&\oln{\Lambda}_{0\pl} = 2 a_0 a_\pl \cos(\Delta_0 - \Delta_\pl) \big( 1 + \e_3 + \e_1\c2b + \e_2\s2b \big)~, \nonumber \\
&\oln{\Lambda}_{\perp i} = 2 a_\perp a_i \big( (\e_2\c2b - \e_1\s2b)\cos\Delta_i + \e_3\sin\Delta_i \big)~, \\
\nonumber\\
\label{etabar}
&\oln{\eta}_{ii} = a_i^2 \big( \e_1 + \c2b +\e_3\c2b \big)~, \nonumber  \\
&\oln{\eta}_{\perp\perp} = a_\perp^2 \big( \e_1 - \c2b - \e_3\c2b \big)~, \nonumber \\
&\oln{\eta}_{0\pl} = 2 a_0 a_\pl \cos(\Delta_0 - \Delta_\pl) \big( \e_1 + \c2b + \e_3 \c2b \big)~, \nonumber \\
&\oln{\eta}_{\perp i} = -2 a_\perp a_i \big( (1 + \e_3)\s2b\cos\Delta_i + \e_1\sin\Delta_i \big)~, \\
\nonumber\\
\label{Sigmabar}
&\oln{\Sigma}_{ii} = -a_i^2 \big( \e_3 + \e_1\c2b + \e_2\s2b \big)~, \nonumber \\
&\oln{\Sigma}_{\perp\perp} = -a_\perp^2 \big( \e_3 - \e_1\c2b - \e_2\s2b \big)~, \nonumber \\
&\oln{\Sigma}_{0\pl} = -2 a_0 a_\pl \cos(\Delta_0-\Delta_\pl) \big( \e_3 + \e_1\c2b + \e_2\s2b \big)~,  \nonumber \\
&\oln{\Sigma}_{\perp i} = -2 a_\perp a_i \big( (\e_2\c2b - \e_1\s2b)\cos\Delta_i +(1 + \e_3)\sin\Delta_i \big)~, \\
\nonumber\\
\label{rhobar}
&\oln{\rho}_{ii} = - a_i^2 \big( \e_2 + \s2b + \e_3\s2b \big)~, \nonumber \\
&\oln{\rho}_{\perp\perp} = - a_\perp^2 \big( \e_2 - \s2b - \e_3\s2b \big)~, \nonumber \\
&\oln{\rho}_{0\pl} = -2 a_0 a_\pl \cos(\Delta_0 - \Delta_\pl) \big( \e_2 + \s2b + \e_3\s2b \big)~, \nonumber \\
&\oln{\rho}_{\perp i} = -2 a_\perp a_i \big( (1 + \e_3)\c2b\cos\Delta_i - \e_2\sin\Delta_i \big) ,
\end{align}
with $\lambda\in\{0,\pl,\perp\}$ and $i \in \{0, \pl\}$.

The solutions for 9 unknown parameters (i.e. 3 of $a_\lambda$, 3 of $\epsilon_j$, 2 of $\Delta_i$ and $\beta$) in terms of observables of $\bar{P}^0\to V_1V_2$ mode are given by:
\begin{align}
\label{alambdaSol2}
&a_\lambda^{}=\sqrt{\oln{\Lambda}_{\lambda\lambda}+\oln{\Sigma}_{\lambda\lambda}}~, \\
\label{eps1Sol2}
& \e_1 = \frac{1}{2}\Big(\frac{\oln{\eta}_{ ii}^{}}{\oln{\Lambda}_{ ii}+\oln{\Sigma}_{ ii}}+\frac{\oln{\eta}_{\perp\perp}^{}}{\oln{\Lambda}_{\perp\perp}+\oln{\Sigma}_{\perp\perp}}\Big)~, \\
\label{eps2Sol2}
& \e_2 = -\frac{1}{2}\Big(\frac{\oln{\rho}_{ ii}^{}}{\oln{\Lambda}_{ ii}+\oln{\Sigma}_{ ii}}+\frac{\oln{\rho}_{\perp\perp}^{}}{\oln{\Lambda}_{\perp\perp}+\oln{\Sigma}_{\perp\perp}}\Big)~, \\
\label{eps3Sol2}
& \e_3 = -\frac{1}{2}\Big(\frac{\oln{\Sigma}_{ ii}^{}}{\oln{\Lambda}_{ ii}+\oln{\Sigma}_{ ii}}+\frac{\oln{\Sigma}_{\perp\perp}^{}}{\oln{\Lambda}_{\perp\perp}+\oln{\Sigma}_{\perp\perp}}\Big)~, \\
\label{SinPetaSol2}
&\sin2\beta=
-\frac{1}{2}\Big(\frac{\oln{\rho}_{ii}}{\oln{\Lambda}_{ii}}-\frac{\oln{\rho}_{\perp\perp}}{\oln{\Lambda}_{\perp\perp}}\Big)~,\\
\label{CosPetaSol2}
&\cos2\beta=
\frac{1}{2}\Big(\frac{\oln{\eta}_{ii}^{}}{\oln{\Lambda}_{ii}}-\frac{\oln{\eta}^{}_{\perp\perp}}{\oln{\Lambda}_{\perp\perp}}\Big)~,\\
\label{CosDeltaDeltaSol2}
&\cos(\Delta_0-\Delta_\parallel)=\frac{1}{2}\Big[\frac{\oln{\Lambda}_{0\parallel}+\oln{\Sigma}_{0\parallel}}{\sqrt{\oln{\Lambda}_{00}+\oln{\Sigma}_{00}}\sqrt{\oln{\Lambda}_{\parallel\parallel}+\oln{\Sigma}_{\parallel\parallel}}}\Big]~,\\
\label{SinDeltaSol2}
& \sin\Delta_i =- \frac{1}{2}\Big[\frac{\oln{\Lambda}_{\perp i}+\oln{\Sigma}_{\perp i}}{\sqrt{\oln{\Lambda}_{ii}+\oln{\Sigma}_{ii}}\sqrt{\oln{\Lambda}_{\perp\perp}+\oln{\Sigma}_{\perp\perp}}}\Big]~, \\
\label{CosDeltaSol2}
&\cos\Delta_i=\oln{X}_i\, \oln{\Lambda}_{ ii}\,\oln{\Lambda}_{\perp\perp}\Big[\frac{\sqrt{\oln{\Lambda}_{ii}+\oln{\Sigma}_{ii}}\sqrt{\oln{\Lambda}_{\perp\perp}+\oln{\Sigma}_{\perp\perp}}}{\oln{\Lambda}_{\perp\perp}\oln{\Sigma}_{ ii}+\oln{\Lambda}_{ii}\oln{\Sigma}_{\perp\perp}+2\oln{\Lambda}_{\perp\perp}\oln{\Lambda}_{ii}}\Big]~, 
\end{align}
\begin{align}
\hspace*{-0.3cm}\text{where, }
\oln{X}_i=\Big[\frac{\big(\oln{\Lambda}_{\perp i}-\oln{\Sigma}_{\perp i}\big)\big(\oln{\Lambda}_{\perp\perp}\oln{\Sigma}_{ ii}+\oln{\Lambda}_{ ii}\oln{\Sigma}_{\perp\perp}\big)+2\big(\oln{\Lambda}_{ ii}\oln{\Lambda}_{\perp\perp}\oln{\Lambda}_{\perp i}-\oln{\Sigma}_{ ii}\oln{\Sigma}_{\perp\perp}\oln{\Sigma}_{\perp i}\big)}{\big(\oln{\eta}_{\perp\perp}^{} \oln{\rho}_{ii}^{}-\oln{\eta}_{ii}^{} \oln{\rho}_{\perp\perp}^{}\big)\big(\oln{\Lambda}_{ii}+\oln{\Sigma}_{ii}\big)\big(\oln{\Lambda}_{\perp\perp}+\oln{\Sigma}_{\perp\perp}\big)}\Big],
\end{align}
with $\lambda\in\{0,\pl,\perp\}$ and $i \in \{0, \pl\}$.

\end{document}